\documentclass[prd,amsmath,amssymb,twocolumn,superscriptaddress,reprint,floatfix,aps]{revtex4-2}
\usepackage{graphicx}
\usepackage{dcolumn}
\usepackage{bm}
\usepackage[utf8]{inputenc}
\usepackage[T1]{fontenc}
\usepackage{mathptmx}
\usepackage{etoolbox}
\usepackage{verbatim}
\usepackage{xcolor}
\usepackage{ulem}

\usepackage{hyperref}% add hypertext capabilities
\hypersetup{colorlinks,%,%
 linkcolor=blue,%
 citecolor=blue,%
 urlcolor=blue
}
\newcommand{\mbr}{\mathbf{r}}
\newcommand{\mbx}{\mathbf{x}}

\newcommand{\mbu}{\mathbf{u}}
\DeclareMathOperator{\sinc}{sinc}
\begin{document}

%\title{Quantum Turbulence in merged self-gravitating Bose-Einstein condensates}
%\title{Quantum Turbulence and Energy Cascades in Merging Self-Gravitating Bose-Einstein Condensates}
%\title{Merging of self-gravitating condensates reveals turbulent dark matter}

\title{Revealing turbulent Dark Matter via merging of self-Gravitating condensates}

\author{Anirudh Sivakumar}%
\affiliation{Department of Physics, Bharathidasan University, Tiruchirappalli 620 024, Tamil Nadu, India}

\author{Pankaj Kumar Mishra}
%\email{pankaj.mishra@iitg.ac.in}
\affiliation{Department of Physics, Indian Institute of Technology Guwahati, Guwahati 781039, Assam, India}

\author{Ahmad A. Hujeirat}
\affiliation{Interdisciplinary Center for Scientific Computing, The University of Heidelberg, 69120 Heidelberg, Germany}

\author{Paulsamy Muruganandam}
%\email{anand@bdu.ac.in}
\affiliation{Department of Physics, Bharathidasan University, Tiruchirappalli 620 024, Tamil Nadu, India}
\affiliation{Department of Medical Physics, Bharathidasan University, Tiruchirappalli 620 024, Tamil Nadu, India}

\begin{abstract}
Self-gravitating condensates have been proposed as potential candidates for modelling dark matter. In this paper, we numerically investigate the dynamics of dark matter utilizing the merging of self-gravitating condensates. We have used the Gross-Pitaevskii-Poisson model and identified distinct turbulent regimes based on the merging speed of the condensate. As a result of collision, we notice the appearance of various dark soliton-mediated instabilities that finally lead to the turbulent state characterized by Kolmogorov-like turbulence scaling \( \varepsilon_{\mathrm{kin}}^i \sim k^{-5/3} \) in the infrared and \( \varepsilon_{\mathrm{kin}}^i \sim k^{-3} \) in the ultraviolet regions. The compressible spectrum suggests weak-wave turbulence. The turbulent fluctuations in the condensate cease as the vortices formed via soliton decay are expelled to the condensate's periphery, manifested in the transferring of kinetic energy from incompressible and compressible flows to the quantum pressure energy. We also establish the significant role played by the self-gravitating trap in determining the distribution of compressible kinetic energy and the resulting density waves, which differ markedly from those observed in atomic condensates under harmonic confinement. Our study may offer valuable insights into the merging of binary stars and open new avenues for understanding the structure and dynamics of the dark matter through self-gravitating condensate.
\end{abstract}

\date{\today}
\maketitle

\section{Introduction}
\label{sec:intro}
Recent theoretical models suggest that dark matter (DM) could consist of ultralight bosons forming a Bose-Einstein condensate (BEC) \cite{Edmonds2017, Boehmer, Chavanis2011a, Chavanis2012, Chavanis2017, Hui2017, Hui2021, Matos}, a concept that has received valuable support from various cosmological simulations \cite{Schive2014, Matos2001, niemeyer2020small}. BECs offer a promising approach to explaining dark matter, as they can account for galactic rotation curves without the need for traditional dark matter models.  The various approximations of BECs are shown to have counterparts in other dark-matter models. The density and pressure of the condensate under non-relativistic Newtonian potential follow the barotropic equation of state. Also the Thomas-Fermi approximation to the condensate follows the Lane-Emden equation~\cite{Madarassy2013}. The ability to adjust the parameters of the BECs provides flexibility, enabling more detailed exploration of dark matter properties.  The weakly interacting Bosons are modeled using the Gross-Pitaevskii (GP) equation, with BEC exhibiting the superfluid behavior. Considering interactions under the Newtonian gravitational framework, these BECs are described by the GP equation coupled with Poisson's equation for the gravitational field, collectively referred to as the Gross-Pitaevskii-Poisson (GPP) model. The approach has been proven to be quite effective in modeling cold galactic dark matter. Unlike conventional atomic BECs, the self-gravitating BECs do not require external potentials for stabilization, relying instead on balance between the repulsive interatomic interactions and quantum pressure~\cite{Chavanis2011a}. Their behavior differs in many ways, especially in the heightened sensitivity of the frequencies of collective oscillation modes to minor changes in condensate mass~\cite{Giovanazzi2001}. O'Dell et al. and Giovanazzi et al. proposed a scheme for electromagnetically (electromagnetically induced ``Gravity'') generating a self-bound Bose-Einstein condensate with $1/r$ attractive interactions- an analog of a Bose star~\cite{O’Dell2000, Giovanazzi2001}.

Apart from explaining the DM, conventional self-gravitating BECs have been proven to be a valuable tool for modeling neutron superfluids within the interiors of pulsars~\cite{Migdal1959, Warszawski2011, Warszawski2012}. Verma et al.~\cite{Verma2022} took a significant step forward by integrating self-gravitating effects and a crust potential to pin vortices in the BEC models. Carrying forward the idea, Shukla et al. \cite{Shukla2024} made further advancements by coupling self-gravitating neutron superfluids with proton superconductors in the presence of magnetic fields. This approach has been extended to explore complex phenomena, such as axionic stars, by incorporating a quintic term in the potential, and is crucial for understanding the dynamics of collapsing stars~\cite{Shukla2024Axion}.

Collisions of self-gravitating BECs have also garnered interest among researchers as galactic collisions, composed of dark and luminous matter, could provide information on the nature of dark matter. During such collisions, luminous matter is expelled from the gravitational potential while simultaneously generating gravitational waves~\cite{Guzman2016}. Similar to the mixing of conventional BECs producing phase-dependent interference patterns~\cite{Xiong2013, Andrews1997, Bloch, Jo2007}, mergers of self-gravitating BECs have also been found to display the similar types of phenomena~\cite{Paredes2016}. It was demonstrated that the collisions between solitonic BEC-dark matter cores with opposite phases yield a short-range repulsive interaction due to destructive interference. Another sort of similarity was shown to exist due to density oscillations in the merged condensate, shown by Scott et al. \cite{Scott} in atomic BECs, which was also complemented by Schwabe \cite{Schwabe2016} in self-gravitating systems.  These self-gravitating BECs have been shown to survive and retain their structural integrity post-collision/merger~\cite{Choi2002, Cotner2016}. In a compelling study, Nikolaieva et al.~\cite{Nikolaieva2023}, utilizing the Gross-Pitaevskii-Poisson model, explored the fascinating collisions of solitonic BECs with vortex structures. Their findings reveal the dominance of the superfluid nature of bosonic dark matter characterized by quantized vortex lines and vortex rings, which manifest in intricate interference patterns during these collisions. The study also highlights that these interference patterns significantly affect the emission of gravitational waves, reinforcing earlier conclusions that vortex structures maintain their integrity even after head-on collisions. Asakawa and Tsubota~\cite{Asakawa2024} explored the dynamics of co-rotating vortices in three-dimensional self-gravitating BECs, analyzing the relationship between vortex separation, their orbital periods and the drag effect exerted by gravitational interactions on quantized vortices. Recently, Asakawa et al. investigated the breathing and anisotropic collective modes of self-gravitating BECs by analyzing their excitations~\cite{Asakawa2024a}.

In recent years, numerical simulations have played a crucial role in uncovering many fascinating features of quantum turbulence in 2D~\cite{Numasato2009, Numasato2009a, Numasato2010, Das2022} and 3D~\cite{Mueller2020, amette2022thermalized, AmetteEstrada2022, makinen2023rotating} atomic BECs. Recently, rotating quantum turbulence has gained significant attraction due to its fascinating properties compared to its non-rotating counterparts \cite{amette2022thermalized, AmetteEstrada2022, makinen2023rotating, sivakumar2023energy, sivakumar2024dynamic}. In most cases, the incompressible kinetic energy spectra exhibit a scaling law consistent with \( E(k) \sim k^{-5/3} \), which is characteristic of vortex cascades. In contrast, the compressible spectra show significant thermalization. An alternative type of quantum turbulence regime, known as Vinen turbulence, is also reported \cite{Cidrim2017, Marino2021} which displays a $E(k) \sim k^{-1}$ scaling, indicating condensates dominated by isolated vortex structures. Another novel regime of quantum turbulence described in the literature called strong quantum turbulence~\cite{barenghi2023types} is shown to arise in condensates with oscillating traps~\cite{middleton2023}, rotating perturbed barriers~\cite{sivakumar2023energy} or even rotational condensate mergers~\cite{sivakumar2024dynamic}. 

Compared to the advances related to turbulence in conventional BECs, the nature of quantum turbulence via BEC-DM collisions has hardly been explored. It has been studied only in the fuzzy dark matter regime, where the condensates lack self-interaction~\cite{Mocz2017, Liu2023}. For instance, Mocz et al.~\cite{Mocz2017}  have shown that the energy spectra follow $E(k) \sim k^{-1.1}$, similar to a thermally driven counterflow rather than a $k^{-5/3}$ Kolmogorov scaling. However, there are few studies available that show the impact of rotation in generating many instabilities that may be a helpful clue for generating turbulence in BEC-DM. For instance, it has been shown that the BEC-DMs subjected to rotation undergo significant structural deformations~\cite{RindlerDaller2012}. The stationary vortices formed in rotating condensates have been investigated quite extensively~\cite{Nikolaieva2021, Dmitriev2021}, revealing that multiply charged vortices are generally unstable over cosmological timescales, except for the singly charged vortex state. Kain et al.~\cite{Kain} compared the critical angular frequency for the emergence of a central vortex in conventional BECs with the rotational speeds of galaxies. On the other hand, Zinner et al.~\cite{Zinner2011} examined the potential for vortex-lattice formation in ultralight BEC-DM systems. 
In this paper, we numerically investigate the nature of turbulent flow in self-gravitating, interacting BEC-DM halos by orchestrating collisions between spatially separated halos at velocities below the quasi-elastic limit to facilitate the formation of a unified BEC-DM structure. We have presented a detailed analysis of the resulting energy spectra, providing valuable insights into the onset and underlying conditions of quantum turbulence within these intriguing astrophysical systems.

Our paper is organized as follows. In Sec. \ref{sec:num}, we introduce the dimensionless Gross-Pitaevskii equation, outline our choice of units, and describe the ansatz and methodology employed in our merger simulation. We also detail the spectral calculations necessary for quantifying quantum turbulence. Section \ref{sec:res} presents the results of merging condensates with unit circulation at various collision velocities. In Sec. \ref{sec:res-merge}, We analyze the formation and evolution of topological structures at times just after the collision. Then, we explore the long-term evolution of the condensate in Section \ref{sec:res-phase}. In Sec. \ref{sec:res-turb}, we describe the onset and nature of turbulence by studying the behavior of kinetic energy components and the spectral dynamics during turbulence. Finally, we conclude in Sec.~\ref{sec:conclusion} by summarizing our findings and discussing the implications of our study for understanding dark matter, as well as the potential of self-gravitating Bose-Einstein condensates as candidates for dark matter.

\section{Numerical Model}
\label{sec:num}
We consider a three-dimensional condensate, which dynamics is described using the dimensionless Gross-Pitaevskii-Poisson system of equations. %
\begin{subequations}%
\begin{align}
  \mathrm{i} \frac{\partial \psi}{\partial t} & = \left(-\frac{1}{2}\nabla^2 + \Phi + g \lvert \psi\rvert^2 \right)\psi, \label{eq:gpe}\\ 
  \nabla^2 \Phi & = \lvert \psi\rvert^2. \label{eq:poisson}
\end{align}
\end{subequations}%
where $\psi \equiv \psi(\mathbf r,t)$ and $\Phi \equiv \Phi(\mathbf{r},t)$ are the condensate wave function and the gravitational potential, respectively, with $\mathbf{r} \equiv (x,y,z)$. $\nabla^2$ is the three-dimensional Laplace operator defined as $\nabla^2 = \partial_x^2 +\partial_y^2 +\partial_z^2$. $g = 4\pi\hbar^2 a_s/m$ is the coupling strength with $a_s$ being the $s$ wave scattering length and $m$ being mass of the particle. We obtain this dimensionless form of the dynamical equations by considering the following transformations $\mathbf{r}\to \mathbf{r}/L$, $t \to t / T$, $E\to E/\epsilon$ of length, time and energy, respectively. Following the formalism described in Ref.~\cite{Nikolaieva2021}, we consider the scaling factor as $L = \lambda_C(m_{\mathrm{pl}}/m)\sqrt{\lambda/8\pi}$, $T = L^2/(c\lambda_C)$ and $\epsilon = (\hbar^2/4\pi m_{\mathrm{pl}}\lambda_C^2) (8\pi/\lambda)^{3/2}$. Here, $m_{\mathrm{pl}} = \sqrt{\hbar c / G}$ is the Planck mass, with $c$ and $G$ being the speed of light and Gravitational constant, respectively. $\lambda_C = \hbar / mc$ is the Compton wavelength for the given particle of mass $m$ and $\lambda / 8\pi = a_s / \lambda_C$ is the scaled scattering length. These transformations also allow one to scale the interaction strength as $g=1$. This normalization condition is given as 
\begin{align}
  \int \lvert\psi\rvert^2 dr = 4\pi \frac{M}{m_{\mathrm{pl}}}\sqrt{\frac{\lambda}{8\pi}} = N_0
\end{align}
For all our simulation runs, we choose the particle mass to be $3\times 10^{-24} \mathrm{eV}$ and the self-interaction constant $\lambda/8\pi = 5.62 \times 10^{-98}$. Fixing these values allows us to tune the total halo mass ($M$) by fixing the number of atoms $N_0$. In our study, we choose $N_0 = 625$ for sustaining several vortices, which results in the total DM halo mass as $M = 2.3 \times 10^{12}M_{\odot}$. These also give us the liberty to estimate the numerical value of scaling factors $L = 6.35\times 10^{19} \mathrm{m}$, $T = 2.04 \times 10^{14} \mathrm{s}$, $\epsilon = 7.05 \times 10^{50} \mathrm{J}$ \cite{Nikolaieva2021}.

The energy components of the Eq.~(\eqref{eq:gpe}) are taken in dimensionless form as%
\begin{subequations}
    \begin{align}
        E_{\mathrm{kin}} & = \frac{1}{2}\int \lvert \nabla\psi\rvert^2 dr, \label{eq:ekin} \\
        E_{\mathrm{gpot}} & = \frac{1}{2}\int \lvert\psi\rvert^2\Phi dr, \label{eq:egravpot} \\
        E_{\mathrm{int}} & = \frac{1}{2}\int \lvert\psi\rvert^4 dr, \label{eq:einteract}
    \end{align}
    \label{eq:enercomp}
\end{subequations}%
where $E_{\mathrm{kin}}$, $E_{\mathrm{gpot}}$, and $E_{\mathrm{int}}$ are the kinetic, potential and interaction energies, respectively, and their sum gives the total energy of the system $E$.

We consider an initial ansatz for the condensate wave function with cylindrical symmetry of the following form~\cite{Nikolaieva2021}:
\begin{align}
    \psi(r_{\perp},z,\theta) = A \left(\frac{r_{\perp}}{R}\right)^s \mathrm e^{-\frac{r_{\perp}^{2}}{2R^2} - \frac{z^2}{2 (R \eta)^2}} \mathrm e^{\mathrm i s\theta},
    \label{eq:ansatz}
\end{align}
where $r_{\perp} = \sqrt{x^2 + y^2}$ is the transverse radius and $\theta = \tan^{-1}(y/x)$.  $R,\eta$ are the variational parameters and $s$ represents the vortex circulation. The factor $A$ is obtained by the normalization condition as $A = \sqrt{N_0 / (\pi^{3/2} \eta R^3 s!)}$.  For all our numerical simulations, we avoid initial conditions with circulation \( s > 1 \) as they are unstable under real-time propagation~\cite{Nikolaieva2021}. Following this approach, we substitute the ansatz (\ref{eq:ansatz}) into Eq.~(\eqref{eq:enercomp}) to compute the total energy \( E \). We then minimize the total energy with respect to the variational parameters \( R \) (transverse condensate width) and \( \eta \) (axial aspect ratio) by considering the equations \( \partial E / \partial R = 0 \) and \( \partial E / \partial \eta = 0 \), which yields the optimal values of these parameters for the given value of \( s \).

For the real-time simulation used to explore the dynamics, we consider an initial wave function of a spatially separated condensate and multiply the wave function of each condensate with a velocity phase exponent, as described in \cite{Nikolaieva2023}, given by:
\begin{align}
    \psi = \psi_{l}\cdot \mathrm e^{-\mathrm i vz} + \psi_{r} \cdot \mathrm e^{\mathrm ivz}
\end{align}
where $\psi_l$ and $\psi_r$ represent the condensates corresponding to the left and right sides of the computational domain, respectively, each having the form described in Eq. \eqref{eq:ansatz}, where \( v \) denotes the translational velocity. The boundary conditions for the potential \( \Phi \) have been implemented by considering both the monopole and quadrupole expansions of \( \Phi \), as follows:
\begin{align}
    \Phi = -\frac{N_0}{4\pi r} - \frac{1}{8\pi r^5}\sum_{i,j}I_{ij}x_i x_j,
    \label{eq:quadpole}
\end{align}
where the quadrupole moment $I_{ij}$ is given by 
\begin{align}
    I_{ij} = \int \lvert\psi\rvert^2\left(x_i x_j - \frac{1}{3}\delta_{i,j} r^2 \right) d\mbr,
\end{align}
with $r = \sqrt{x^2 + y^2 + z^2}$, $\delta_{i,j}$ is the Kroenecker delta function. Note that we have used a density-dependent boundary condition, which requires updating at each time step.

To characterize the turbulence and energy cascades resulting from vortex dynamics, we perform a detailed analysis of the kinetic energy spectra across various spatial and temporal scales. We decompose the kinetic energy into its compressible and incompressible components \cite{Nore1997} by transforming the respective velocity fields into $k$-space integrals using Parseval's theorem. For this purpose, we utilize the numerical implementation developed by Bradley et al.~\cite{Bradley2022}, which calculates these integrals using the angle-averaged Wiener-Khinchin theorem, as demonstrated below for the incompressible kinetic energy spectrum.
\begin{align}
    \varepsilon_{\mathrm{kin}}^i (k) = \frac{m}{2}\int d^d\mathbf{x} \, \Lambda_d\left(k, \lvert \mathbf{x} \rvert\right)\, C\left[\mathbf{u}^i, \mathbf{u}^i\right](\mathbf{x}),
    \label{eq:wkspec}
\end{align}
where \(\varepsilon_{\mathrm{kin}}^i(k)\) is the angle-averaged incompressible kinetic energy spectrum, \(d\) specifies the spatial dimension. \(C[\mbu^i, \mbu^i](\mbx)\) is the two-point auto-correlation function in coordinate (position) space for a given incompressible velocity field, and \(\Lambda_d(k, \lvert\mathbf{x}\rvert)\) is the dimension-dependent kernel function described as 
\begin{align}
  \Lambda_{d}(k,r) = 
  \begin{cases}
    \frac{1}{2\pi}k J_0(kr), & \mbox{for } d=2 , \\
    \frac{1}{2\pi^2}k^2 \sinc(kr), & \mbox{for } d=3. \\
  \end{cases}
  \label{eq:kernel}
\end{align}
Eq.~\eqref{eq:wkspec} implies that for any position-space field, there exists a spectral density equivalent to an angle-averaged two-point correlation in $k$ space.

To further characterize the energy cascades, we consider several length scales: \(R_{\mathrm{TF}}\), \(\ell_0\), \(\xi_c\), and \(\xi\). Here, \(R_{\mathrm{TF}}\) denotes the Thomas-Fermi radius, which for self-gravitating systems is defined as \(R_{\mathrm{TF}} = \pi \sqrt{a_s \hbar^2 / G m^3}\) \cite{Asakawa2024, Zhang2018}. \(\ell_0\) represents the inter-vortex distance, which can be calculated using the relation \(\ell_0 = (V / L_{v})^{1/2}\), where \(V\) is the condensate volume and \(L_{v}\) is the vortex length, is given by~\cite{AmetteEstrada2022}:
\begin{align}
    L_v = 2\pi \frac{\int_{k_{\mathrm{min}}}^{k_{\mathrm{max}}} n(k) \, dk}{\int_{k_{\mathrm{min}}}^{k_{\mathrm{max}}} n_{s}(k) \, dk},
\end{align}
where \(n(k)\) is the momentum spectrum of the condensate, and \(n_s(k)\) is the momentum spectrum of a single-vortex condensate.  \( k_{\mathrm{min}} \) is the minimum value from which we begin integrating the spectra, and we choose \( k_{\mathrm{min.}} = 10 \). \( k_{\mathrm{max.}} \) denotes the maximum value in the \( k \)-array. The parameter \( \xi \) corresponds to the healing length of the condensate, while \( \xi_c = \sqrt{\frac{m}{8\pi\rho_c a_s}} \) is the healing length associated with the critical density \( \rho_c \). It is important to note that vortices with core sizes larger than \( \xi_c \) are energetically unstable and cannot be sustained in the condensate for longer durations \cite{Nikolaieva2021, Dmitriev2021, Asakawa2024}.

\section{Results}

\label{sec:res}
\subsection{Dynamics of dark matter using self-gravitating Bose-Einstein condensate}
\label{sec:res-merge}
Mergers of cosmological dark matter structures have been shown to exhibit nonlinear behavior and significantly affect the nature of the resulting dark matter halo~\cite{Carbone2006, Quilis, Quilis1998}. In addition, the formation of topological defects can alter the galactic rotation curves and the radiation of gravitational waves, which could be detected directly due to their sensitivity to dark matter distribution. Even though several numerical models of dark matter densities are present, in this paper we utilize an energetically favorable initial density profile, based on the mean-field model which provides an interesting dynamical evolution of dark matter distributions.

We begin our analysis by analyzing the fascinating collision of two spatially separated condensates, each featuring a single central vortex with a circulation of \( s = 1 \). We explore the key features of the collisional dynamics by numerically solving the GPE using the Crank-Nicolson scheme \cite{muruganandam2009fortran, kumar2019c, young2017openmp, vudragovic2012c} and solving the Poisson equation using the pseudospectral method in a computational domain of size \( 512^3 \) with step sizes of \( dx = dy = dz = 0.1 \). These condensates are set in motion at distinct velocities, specifically chosen to remain below \( v = 9 \times 10^{-4} c \).%
\begin{figure}[ht!]
    \centering
    \includegraphics[width=\linewidth]{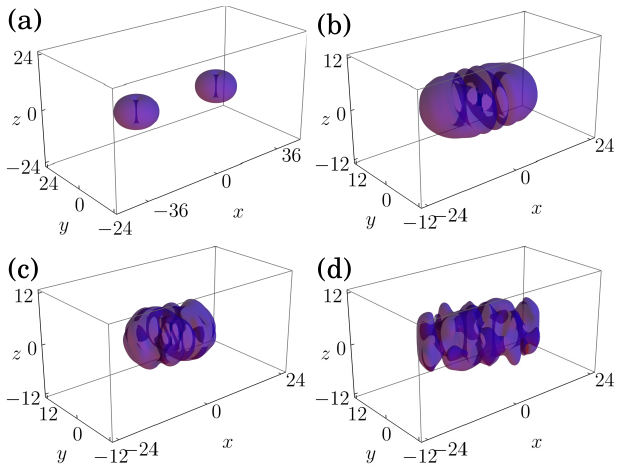}
    \caption{
    Plots showing three-dimensional density contours (isosurfaces) of colliding condensates at a velocity of \( v = 9 \times 10^{-4}c \), depicted at different times to illustrate the merger of self-gravitating BECs: (a) \( t = 0 \), (b) \( t = 32 \times 10^6 \) years, (c) \( t = 38 \times 10^6 \) years, and (d) \( t = 180 \times 10^6 \) years. All contours represent the density level \( \lvert \psi \rvert^2 = 0.1 \).  
    }
    \label{fig:3d}
\end{figure}%
\begin{figure*}[ht!]
    \centering
    \includegraphics[width=\linewidth]{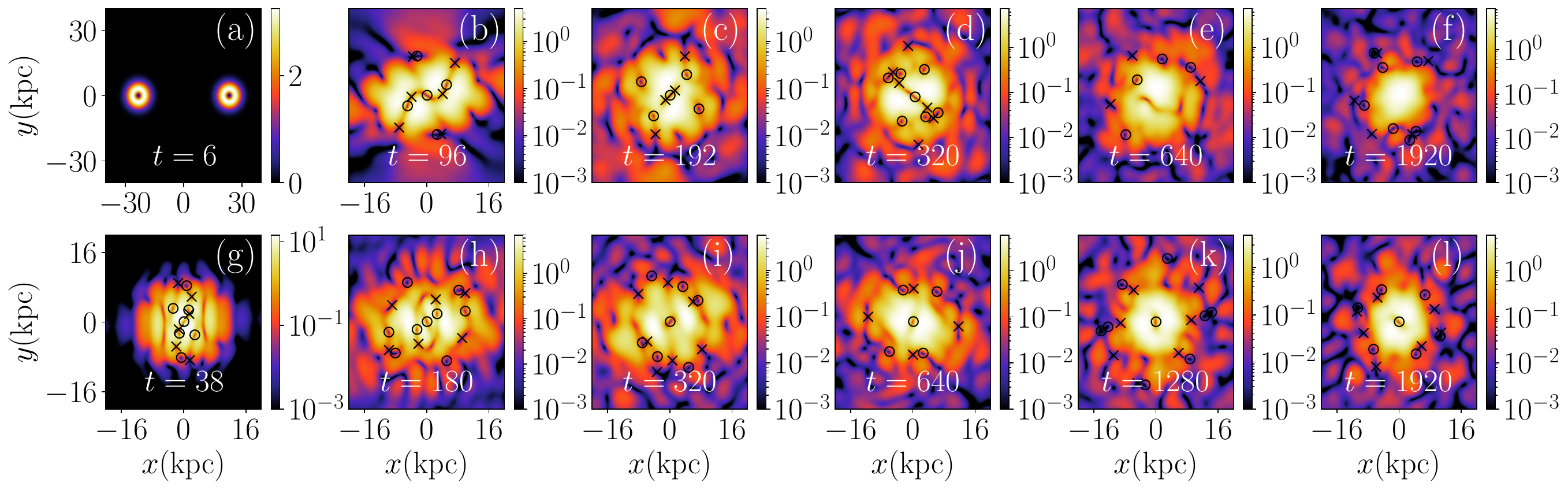}
    \caption{Plots showing the (a) two-dimensional density profile of the initial condensate (b)-(l) log-normalized  density profiles of the merging condensates with circulation \( s = 1 \), along the \( x\text{-}y \) plane (\( z = 0 \)). The evolution of the condensate is illustrated for collisional velocities: (b)-(f) \( v = 6 \times 10^{-4}c \) and (g)-(l) \( v = 9 \times 10^{-4}c \). Time is measured in units of \( 10^6 \) years, and condensate density is expressed in units of \( 2 \times 10^6 \, M_{\odot}/\mathrm{kpc}^3 \), where \( M_{\odot} \) denotes the solar mass. The black circles and crosses mark the positions of vortices and antivortices respectively.
    }
    \label{fig:densityxy_s11}
\end{figure*}
When velocities surpass this critical threshold, the collision between the condensates is elastic, causing the condensates to pass through one another seamlessly. However, at velocities below this limit (\textcolor{blue}{$v < 9\times 10^{-4}c$}), the condensates get merged into a single blob, which further displays interesting non-equilibrium dynamics and tends towards the turbulence state, which exploration is one of the prime focus of our present work. While we have employed a dimensionless equation in our analysis, we include the relevant length and time scales in all our simulation results reported here to enrich our findings, providing a more compelling portrayal of the dynamics in a more realistic context.

Fig.~\ref{fig:3d} shows the condensate merger for velocity $v=9\times 10^{-4}c$, where the isosurfaces  depict the formation of ring solitons (see Figs.~\ref{fig:3d}(b) \& ~\ref{fig:3d}(c)). The bending and twisting of vortex filaments can also be observed in Fig.~\ref{fig:3d}(d).
To efficiently visualize the dynamics, we consider the log-normalized, pseudo-color density plots of the condensate mergers in Fig.~\ref{fig:densityxy_s11}.
These plots show the emergence of interference patterns (see Fig.~\ref{fig:densityxy_s11}(b),(g)). Due to the lower velocity, these fringes are not strong enough to repel the condensates from one another; instead, they coalesce into a single condensate. These interference patterns evolve into dark-ring solitonic structures in the 3D condensate (appearing as dark stripes in the 2D $x-y$ plane [see Fig.~\ref{fig:densityxy_s11}(b)(g) and Fig.~\ref{fig:3d}(b)(c)]. The appearance of such solitons has also been reported in the mergers of atomic BECs \cite{sivakumar2024dynamic, Chikkatur2002, Buchmann2009}. These solitons decay through a process known as Snake instability \cite{Verma2017}, in which they break down into counter-rotating vortex pairs that eventually annihilate one another. Similarly, the solitons formed in our self-gravitating systems decay into vortex antivortex pairs via the Snake instability, eventually leading to the turbulent condensate. 
We notice that for all the collisional velocities considered, the vortices and antivortices formed via soliton are expelled from the condensate, highlighting their transient nature. The only exception is the central vortex, which emerges from the merger of the $s=1$ vortices present in the original condensates. It is also evident that faster collisions lead to the creation of more vortices, predominantly driven by the generation of ring solitons. The vortex-antivortex population eventually dwindles as the pairs annihilate to form density waves or get expelled to the lower density regions of the condensate.

\begin{figure}
    \centering
    \includegraphics[width=\linewidth]{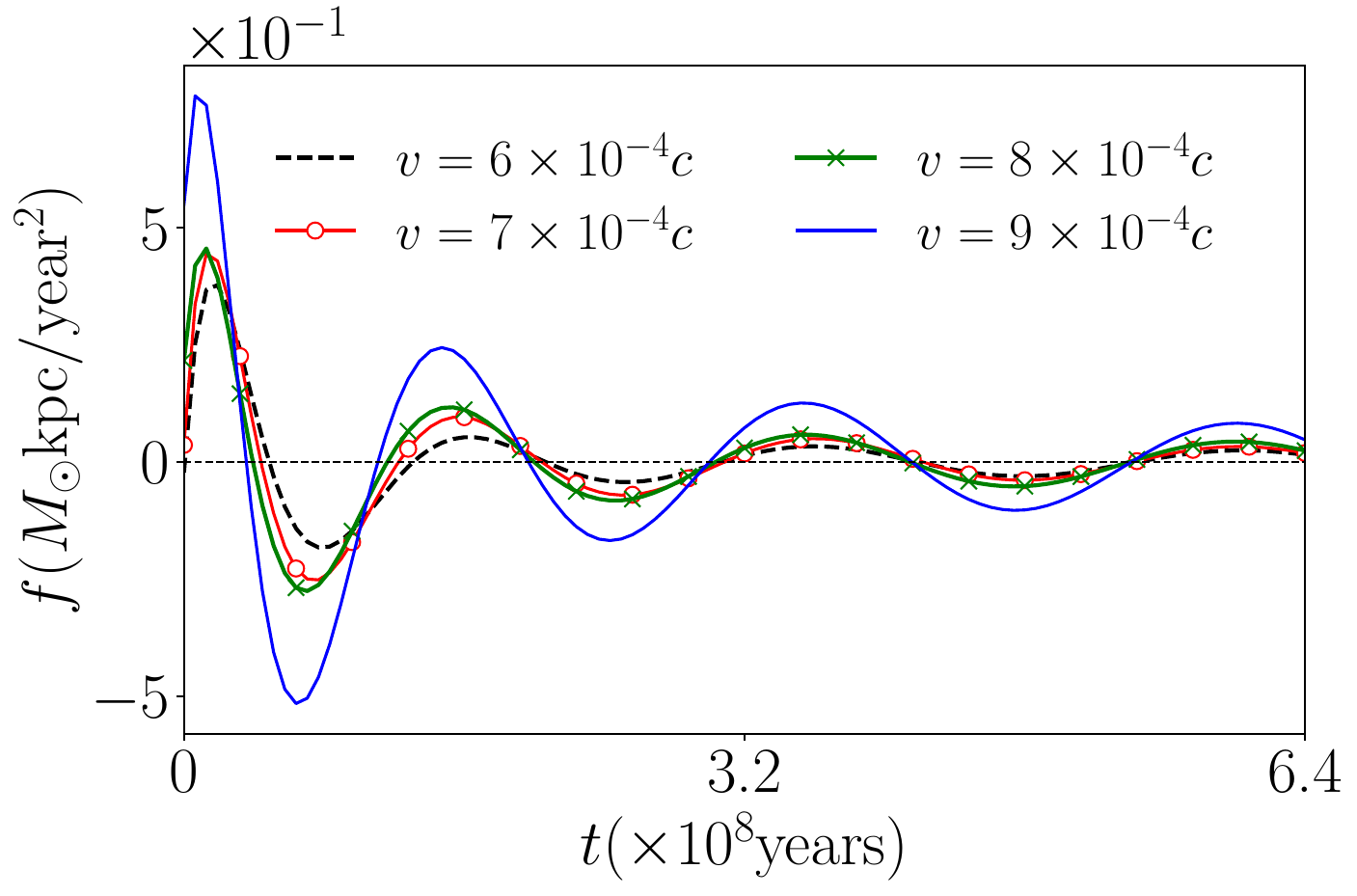}
    \caption{
    Temporal profiles of the total forcing $f$ within the 3D condensate for various collisional velocities. The initial peaks indicate soliton formation, while the subsequent decrease in forcing reflects the timescale of soliton decay via the Snake instability.
    }
    \label{fig:forcing3d}
\end{figure}

The presence of solitons and their nature can be inferred from the total forcing term which is given by 
\begin{align}
f = \frac{d}{dt}\int_{k_{\mathrm{min}}}^{k_{\mathrm{max}}} n(k) dk,
\end{align}
where \(n(k)\) is the 3-D momentum spectrum of the condensate, which we integrate from \(k_{\mathrm{min}} = 10\) to \(k_{\mathrm{max}}\) which is the maximum value of  \(k\) array. During faster collisions, ring solitons are formed with a sharper phase difference across the $x$ axis and are almost stationary. This is indicated by a larger amplitude in the forcing term \( f \) (see Fig.~\ref{fig:forcing3d}). As the solitons eventually propagate with a finite velocity, they decay via Snake instability into vortex-antivortex pairs, decreasing their phase difference and reducing the total forcing within the condensate. The solitons are reflected back and forth and constantly change direction as they propagate across the condensate, and this feature manifests with an appearance of oscillations between positive and negative values in the forcing term.
\begin{figure}[!htb]
    \centering
    \includegraphics[width=0.9\linewidth]{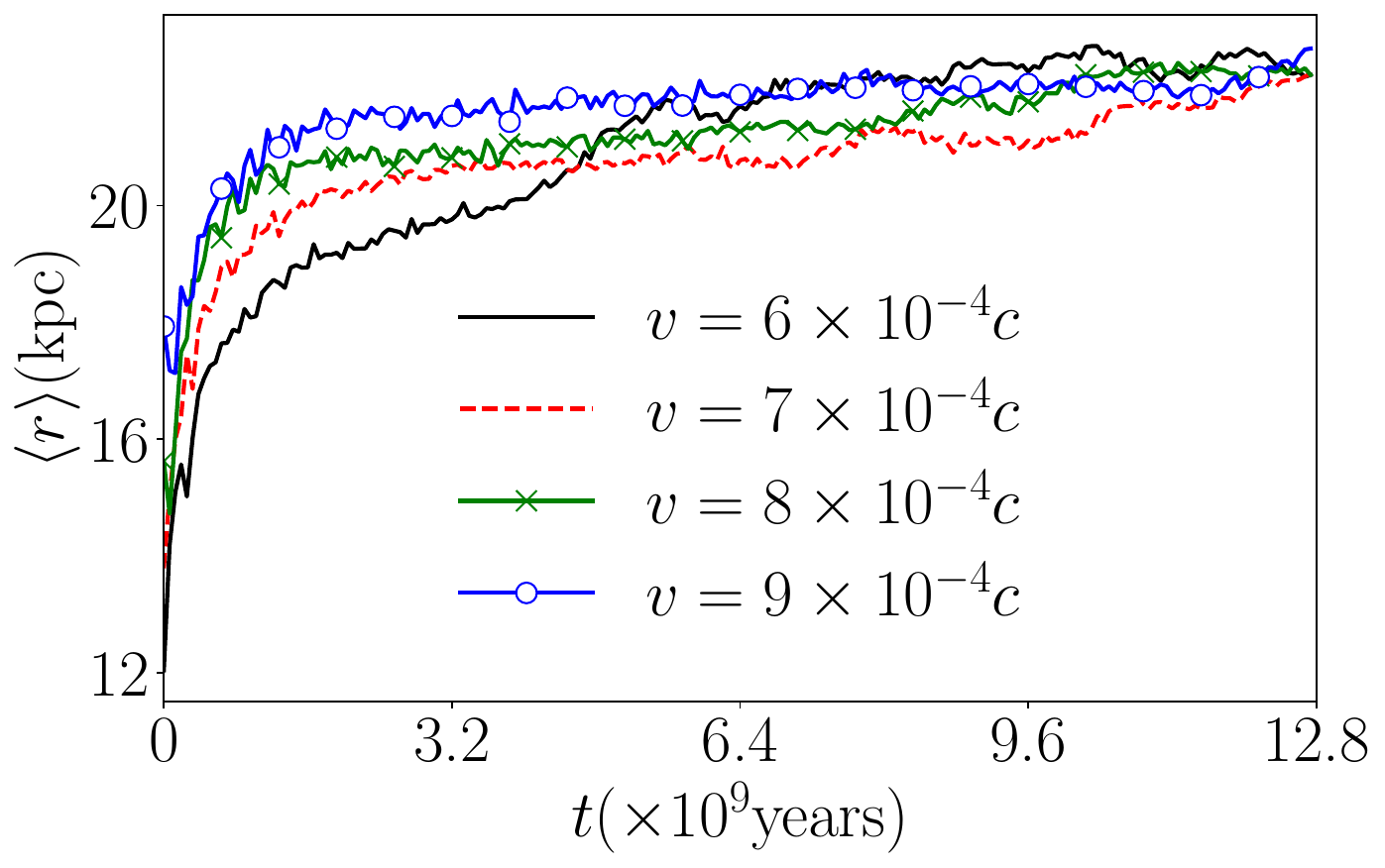}
    \caption{
    Temporal evolution of the 3D condensate radius \(\langle r \rangle\) for different collisional velocities (\(v\)). The onset of vortex collapse, indicated by an increase in the condensate radius, occurs much earlier in slower collisions than in faster ones. The inset shows the radius over shorter time scales (in units of \(10^6\) years).
    }
    \label{fig:radius3d_s11}
\end{figure}

\subsection{Different dynamical phases of Dark Matter due to merging}
\label{sec:res-phase}

In recent times, there have been several studies that highlight the complex dynamical structure of the galactic halo mergers upon varying the velocities, rotation frequencies, and mass distributions and explore their impacts on the gravitational wave signatures and rotation curves \cite{inagaki2010systematic}. These gravitational waves are known to alter the polarization of the cosmic microwave background, which can aid in detecting these collisions and the presence of vortex structures. Following this, in this section, we investigate the different dynamical phases that arise due to the merging of the dark matter condensate that we will show eventually leading to the turbulent state. Further, we will characterize the impact of the collision between the dark matter condensates using gravitational luminosity.

While the short-term dynamics exhibit turbulence, the long-term evolution of the merged condensate reveals intriguing behavior. Notably, density profiles (Fig.~\ref{fig:densityxy_s11}) show that vortices are confined to the outer edges of the condensate, while a central vortex with circulation $s = 2$ persists. This central vortex forms as a result of the combination of two $s = 1$ vortices during the merger. Over time, the $s = 2$ vortex also migrates to the periphery, as demonstrated by Nikolaieva et al.~\cite{Nikolaieva2021}. The decay of the $s = 2$ vortex is strongly influenced by the collision velocity, as reflected in the increasing condensate radius $\langle r \rangle$, illustrated in Fig.~\ref{fig:radius3d_s11}. %
\begin{figure}[!ht]
    \centering
    \includegraphics[width=0.9\linewidth]{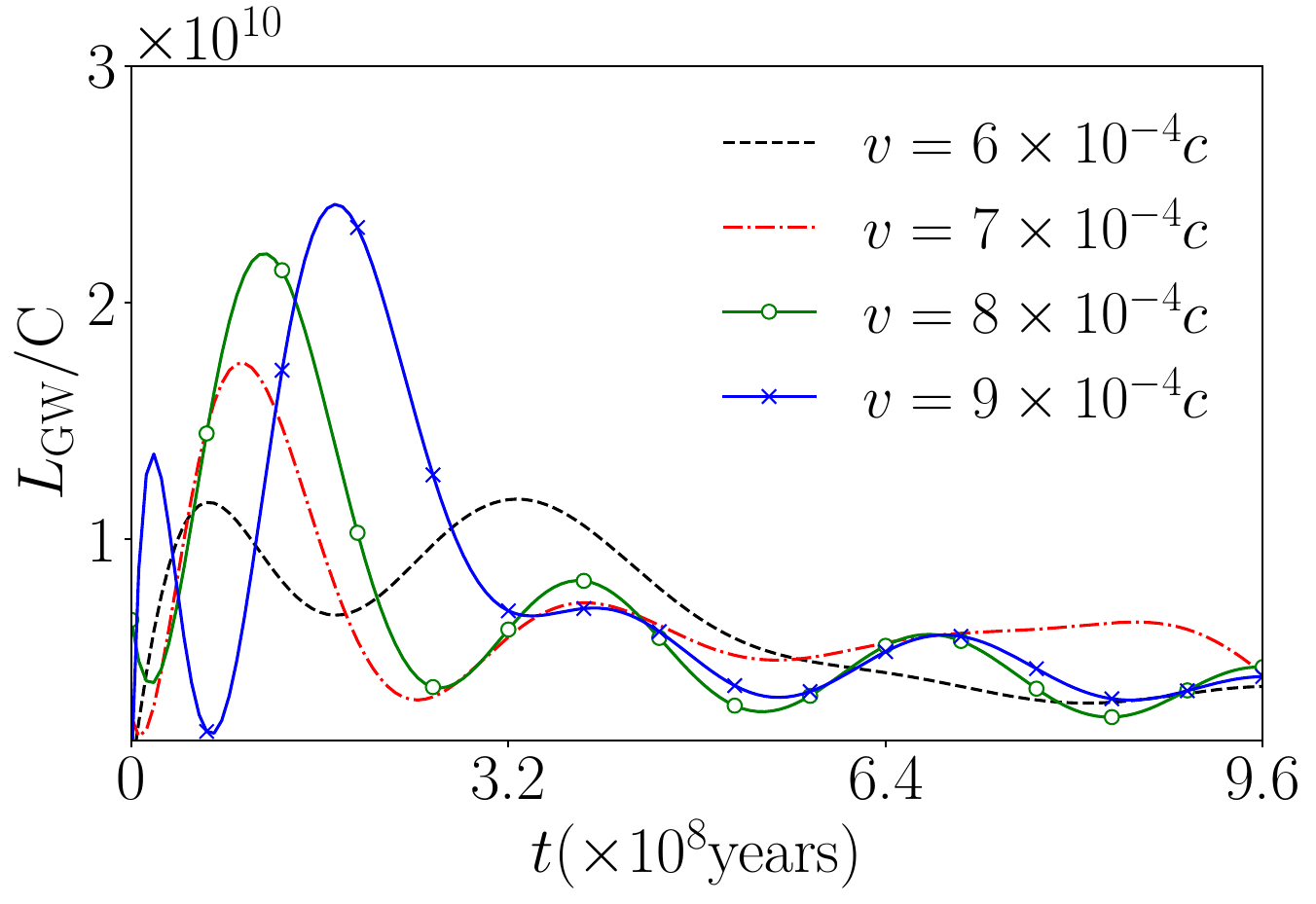}
    \caption{Gravitational wave luminosity profiles for various collision velocities, calculated using all components of the quadrupole moment. Notably, the luminosity profiles of the individual components differ only slightly from the overall profile presented.}
    \label{fig:gw3d_s11}
\end{figure}
Slower mergers lead to a quicker collapse of the vortex structure, suggesting that the condensate with the $s = 2$ vortex forms much earlier at lower collision velocities ($v=6\times10^{-4} c$) and hence collapses within the timeline of the universe as shown by the density profile in Fig.~\ref{fig:densityxy_s11}(f), whereas the $s=2$ vortex persists at the same time frame without collapsing for faster collisions [$v=9\times 10^{-4}c$, see Fig.~\ref{fig:densityxy_s11}(k)-(l)].

We also calculate the luminosity of gravitational waves emitted by the condensate collisions. This luminosity is determined by the third-order time derivative of the quadrupole moment, \( I_{ij} \), and is expressed as \cite{Quilis}:  
\begin{align}  
    L_{\mathrm{GW}} = C\sum_{i,j}\lvert \dddot{I}_{ij} \rvert^2,  
\end{align}  
where \( C = G^4 m^5 / (80\pi^2 c^2 a_s^2) \). %
\begin{figure}[!ht]
    \centering\includegraphics[width=\linewidth]{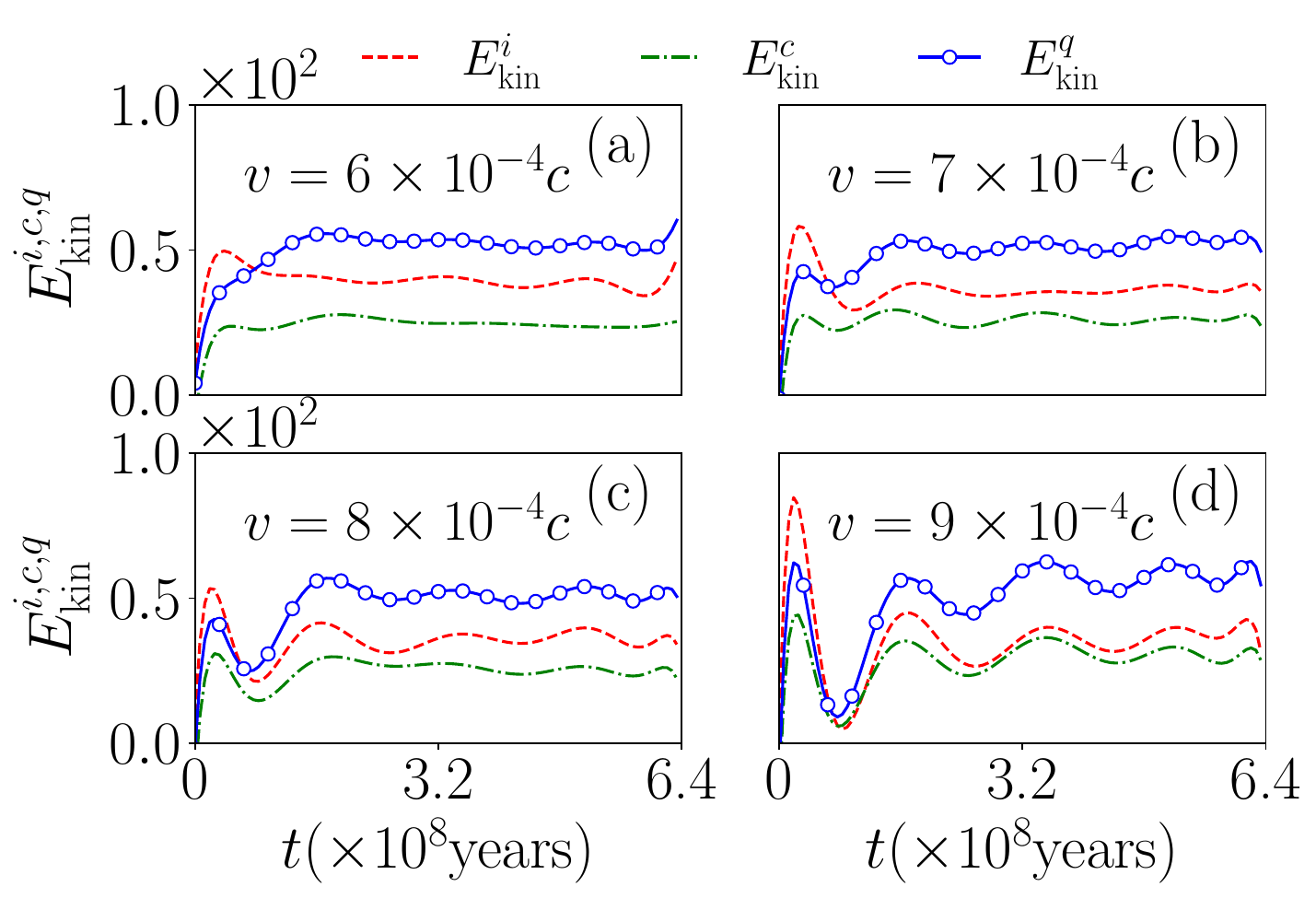}
    \caption{Three dimensional spatial average of the kinetic energy components for different velocities: (a) \(v=6\times10^{-4}c\), (b) \(v=7\times10^{-4}c\), (c) \(v=8\times10^{-4}c\), and (d) \(v=9\times10^{-4}c\). The incompressible kinetic energy (\(E_{\mathrm{kin}}^{i}\)) is represented by red dashed lines, the compressible kinetic energy (\(E_{\mathrm{kin}}^{c}\)) by green dash-dotted lines, and the quantum pressure energy (\(E_{\mathrm{kin}}^{q}\)) by blue circled lines.}
    \label{fig:kincomp3d_s11}
\end{figure}
As pointed out in Ref. \cite{Quilis}, the gravitational wave luminosity is strongly correlated with the amplitude of the gravitational waves themselves. The luminosity profile calculated in the present work is similar to that of head-on collisions of galactic dark matter. It is also important to note that, since the timescale of the oscillations is on the order of $10^8$ years, the associated frequencies are on the order of $\sim 10^{-15} $ Hz, which makes them quite difficult to detect. Similar to the luminosity profiles presented in \cite{Nikolaieva2023}, there is an increase in the amplitude when the condensates overlap, as illustrated in Fig.~\ref{fig:gw3d_s11}.  For lower velocities, the condensates do not exhibit any back-and-forth oscillation due to their low initial momentum, an effect that is also noticeable in the non-oscillatory nature of the gravitational-wave luminosity.
\begin{figure}
    \centering
    \includegraphics[width=\linewidth]{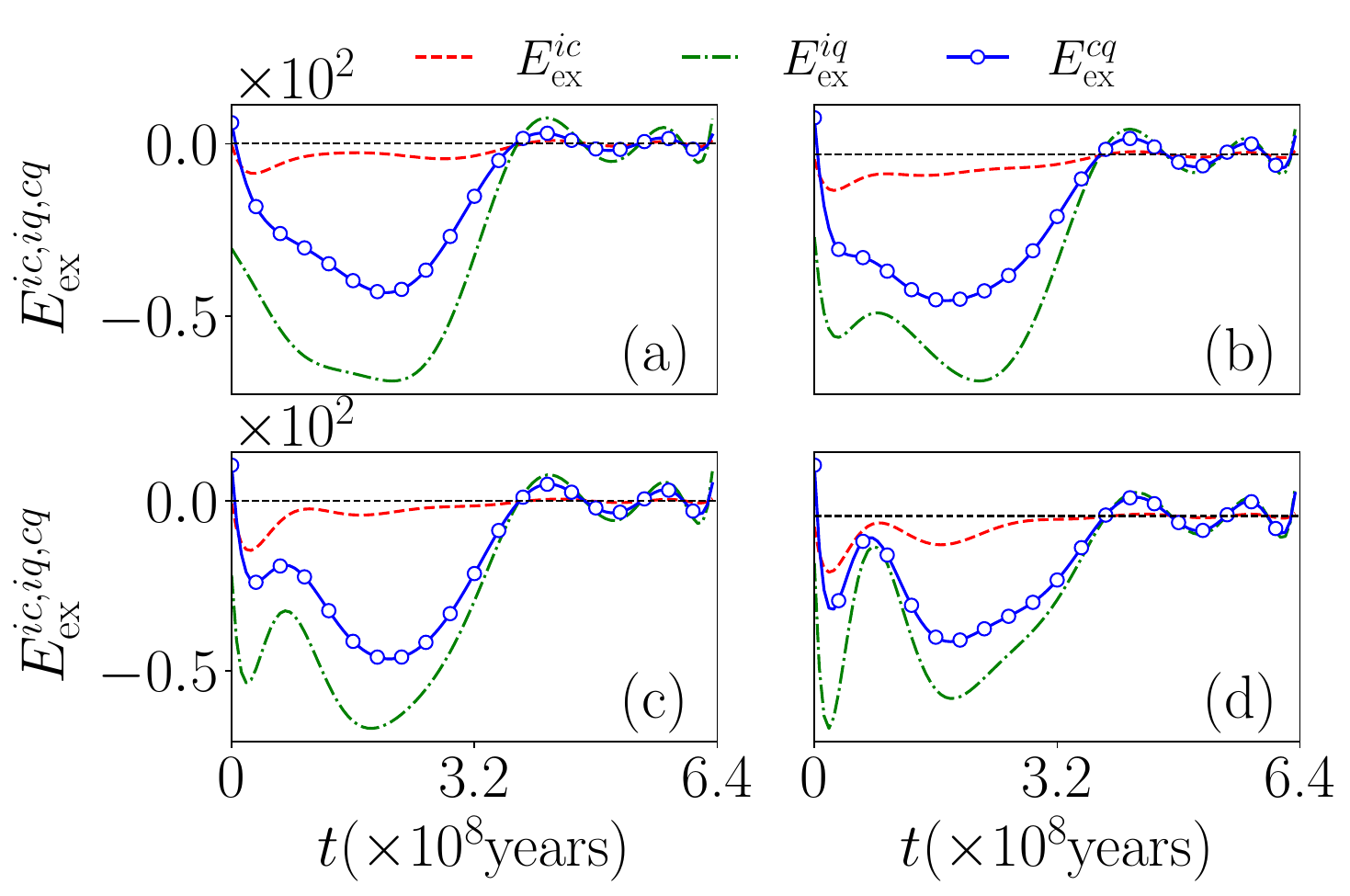}
    \caption{Profiles of total kinetic energy exchange between its components with respect to time for different collisional velocities: (a) \(v=6\times10^{-4}c\), (b) \(v=7\times10^{-4}c\), (c) \(v=8\times10^{-4}c\), and (d) \(v=9\times10^{-4}c\). The red dashed line represents the energy transfer from the incompressible to the compressible component (\(E^{ic}_{\mathrm{ex}}\)), the green dash-dotted line represents the energy transfer between the incompressible and quantum pressure components (\(E^{iq}_{\mathrm{ex}}\)), and the blue line with circles represents the transfer between the compressible and quantum pressure components (\(E^{cq}_{\mathrm{ex}}\)). A negative exchange energy indicates the transfer of energy from the first component in the superscript to the next one, while a positive value indicates the opposite.}
    \label{fig:exchange3d_s11}
\end{figure}
\begin{figure}[!htb]
    \centering
    \includegraphics[width=0.99\linewidth]{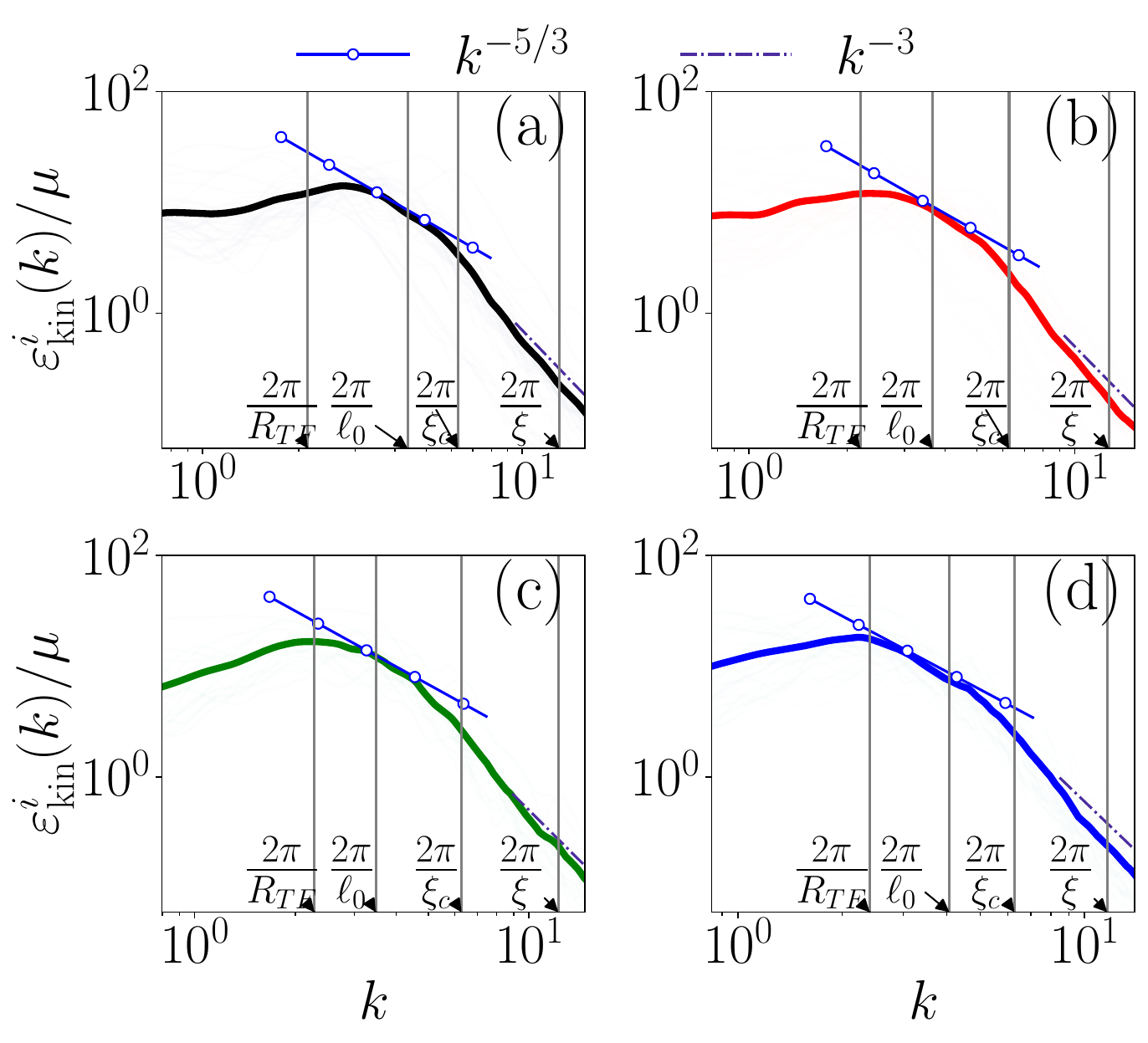}
    \caption{Incompressible kinetic energy spectra for condensates with circulation \(s = 1\) and varying collision velocities: (a) \(v = 6 \times 10^{-4}c\), averaged over the time range \(t = 51\) to \(t = 320\) million years; (b) \(v = 7 \times 10^{-4}c\), in the time range \(t = 51\) to \(t = 320\) million years; (c) \(v = 8 \times 10^{-4}c\), in the time range \(t = 128\) to \(t = 320\) million years; (d) \(v = 9 \times 10^{-4}c\), in the time range \(t = 128\) to \(t = 320\) million years. All cases exhibit \(k^{-5/3}\) scaling in the inertial range and \(k^{-3}\) scaling at larger values of \(k\).}
    \label{fig:ikinspectra3d_s11}
\end{figure}
On the contrary, during faster collisions, due to their large initial momenta, the condensates tend to move away from each other post-collision before returning to the central position, thus overlapping several times even after the initial collision. Hence, the luminosity profile in these cases displays an oscillatory behavior due to multiple low-intensity overlaps. 

\subsection{Characterization of turbulent phase of the Dark Matters}
\label{sec:res-turb}

So far, previous studies resorted to the GPP model to simulate turbulent behavior in mergers of BEC DMs \cite{Mocz2017} and have shown the existence of a $k^{-1.1}$ power law scaling in the kinetic energy spectrum, attributed to a thermally driven counterflow. In this section, in line with the previous studies, we show that the turbulent dark matter condensate undergoes relaxation through the vortex reconnections.

The total kinetic energy within the condensate, illustrated in Fig.~\ref{fig:kincomp3d_s11}, peaks when the condensates overlap with each other. Similar to standard atomic BECs, the incompressible kinetic energy dominates over the compressible component.  However, it is worth noting that the compressible energy is almost comparable to the incompressible component, especially during faster collisions. This behavior is markedly different from that observed in harmonically confined BECs, where the compressible energy is significantly smaller than the incompressible component. Also, the quantum pressure component of the kinetic energy, which doesn't contribute to the velocity field of the condensate, is shown to increase at later times and dominate over the other components. This dominance is explained by the exchange energy density between the components as shown in Fig.~\ref{fig:exchange3d_s11}, where apart from the case where incompressible kinetic energy is converted to its compressible counterpart there is also a transfer of energy from components associated with velocity flow to quantum pressure energy, shown by the negative values of $E_{\mathrm{ex}}^{iq}$ and $E_{\mathrm{ex}}^{cq}$. This corresponds well to the increase of the quantum pressure energy in Fig.~\ref{fig:kincomp3d_s11}, where the quantum pressure component increases initially and saturates at later times when the exchange energy densities approach zero. This suppression of condensate flow via energy exchange within the condensate is the reason the turbulent behavior does not persist and decays relatively quickly. 

It is worth mentioning that the stability of a self-gravitating condensate relies on balancing the attractive potential with the repulsive effects contributed via interatomic interactions (isotropic pressure) and quantum pressure (anisotropic pressure). Based on the analysis from \cite{Nikolaieva2021}, we deduce that for the chosen number of atoms ($N_0$), the energy associated with the interaction pressure is around $10^3$ times the quantum pressure. So, the repulsive interaction term is chiefly responsible for maintaining the condensate stability compared to quantum pressure. The dominance of quantum pressure in later times is not of major significance to condensate stability but is merely a consequence of the suppression of condensate flow.

The vortices formed due to soliton decay can contribute to turbulent fluid flows, as evidenced by the incompressible kinetic energy spectra shown in Fig.~\ref{fig:ikinspectra3d_s11}. The spectral average is taken for time intervals immediately after the collision, revealing a clear \(k^{-5/3}\) Kolmogorov-like scaling for cases with different collisional velocities. This energy cascade is primarily driven by vortices generated through soliton decay, as confirmed by the longer and more pronounced scaling range for faster collisions, which produce more solitons and, consequently, more vortices. The incompressible spectra also demonstrate an enstrophy cascade, highlighting the essential conservation of vorticity within the resulting condensate, characterized by a \( k^{-3} \) power-law scaling at larger \( k \) values. It is crucial to delineate the ranges of these scalings. The \( k^{-5/3} \) scaling terminates at length scales on the order of the critical healing length \( \xi_c \), signifying that vortex energy dissipates into structures with core sizes comparable to \( \xi_c \). Further breakdown of vortices is limited by the core size of roughly on the order of \( \xi \) and determined by the chemical potential \( \mu \). Moreover, the \( k^{-5/3} \) scaling persists on both sides of the inter-vortex distance \( \ell_0 \), reinforcing the presence of a Kolmogorov energy cascade, consistent with previous observations in atomic BECs \cite{sivakumar2023energy, sivakumar2024dynamic}.

\begin{figure}[!htb]
\centering
\includegraphics[width=0.99\linewidth]{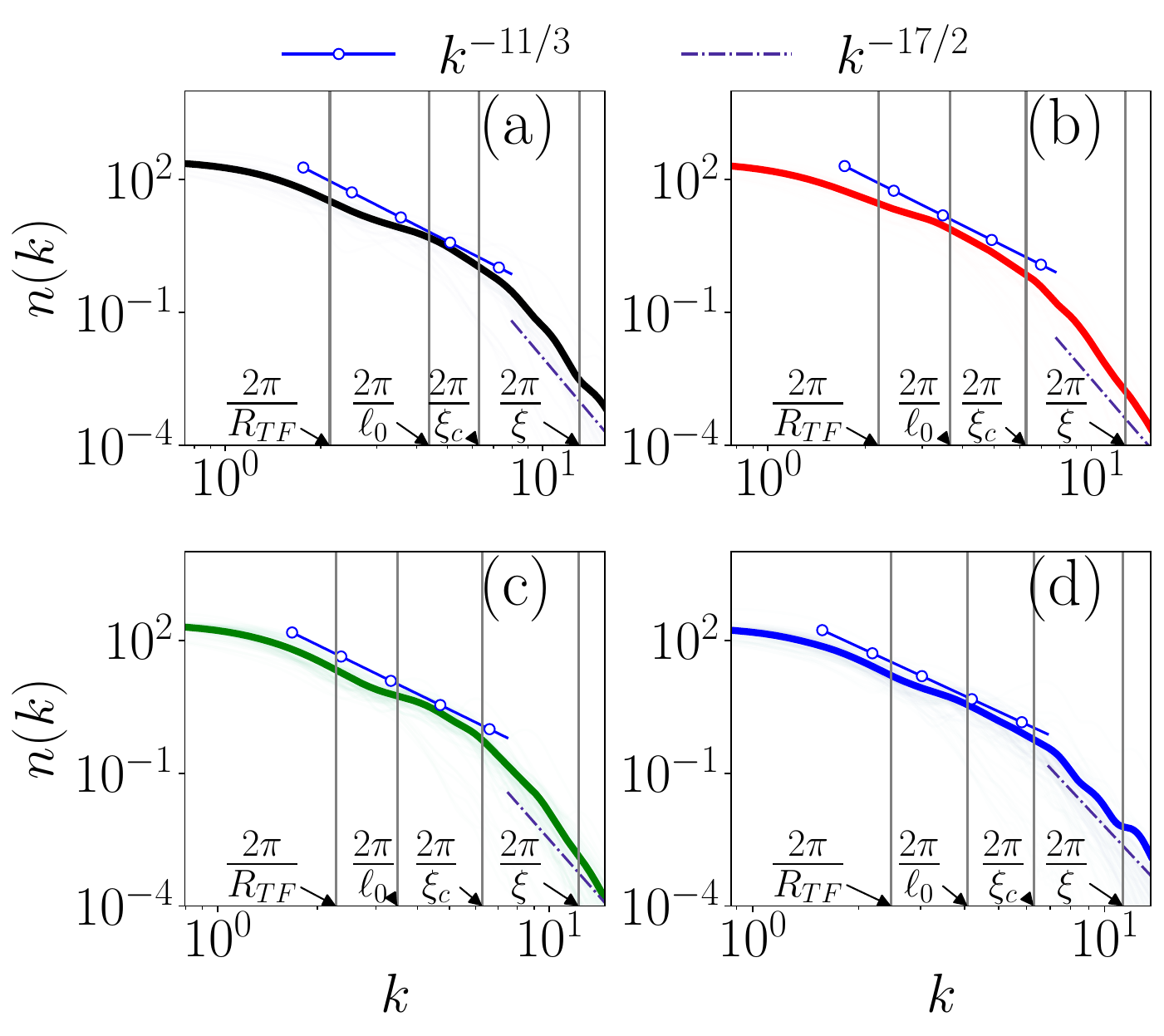}
\caption{Density spectra for condensates with circulation \(s = 1\) at different collision velocities: (a) \(v = 6 \times 10^{-4}c\), averaged over the time range \(t = 51\) to \(t = 320\) million years; (b) \(v = 7 \times 10^{-4}c\), averaged over the same time range; (c) \(v = 8 \times 10^{-4}c\), averaged over \(t = 128\) to \(t = 320\) million years; and (d) \(v = 9 \times 10^{-4}c\), averaged over \(t = 128\) to \(t = 320\) million years. All cases exhibit \(k^{-11/3}\) scaling in the inertial range and \(k^{-17/2}\) scaling at higher values of \(k\).
}
\label{fig:denspectra3d_s11}
\end{figure}

The density spectra of the merged condensate (see Fig.~\ref{fig:denspectra3d_s11}) exhibit a scaling behavior consistent with both the incompressible and compressible spectra. At smaller wave numbers (\(k\) values) on larger scales, the density spectra are primarily influenced by the incompressible kinetic energy spectra. 
This relationship follows a power law of the form \(n(k) \sim \varepsilon_{\mathrm{kin}}^i(k) / k^2\), resulting in a scaling of \(k^{-11/3}\), which corresponds to the \(k^{-5/3}\) scaling found in the incompressible spectra.

\begin{figure}[!htb]
    \centering
    \includegraphics[width=0.99\linewidth]{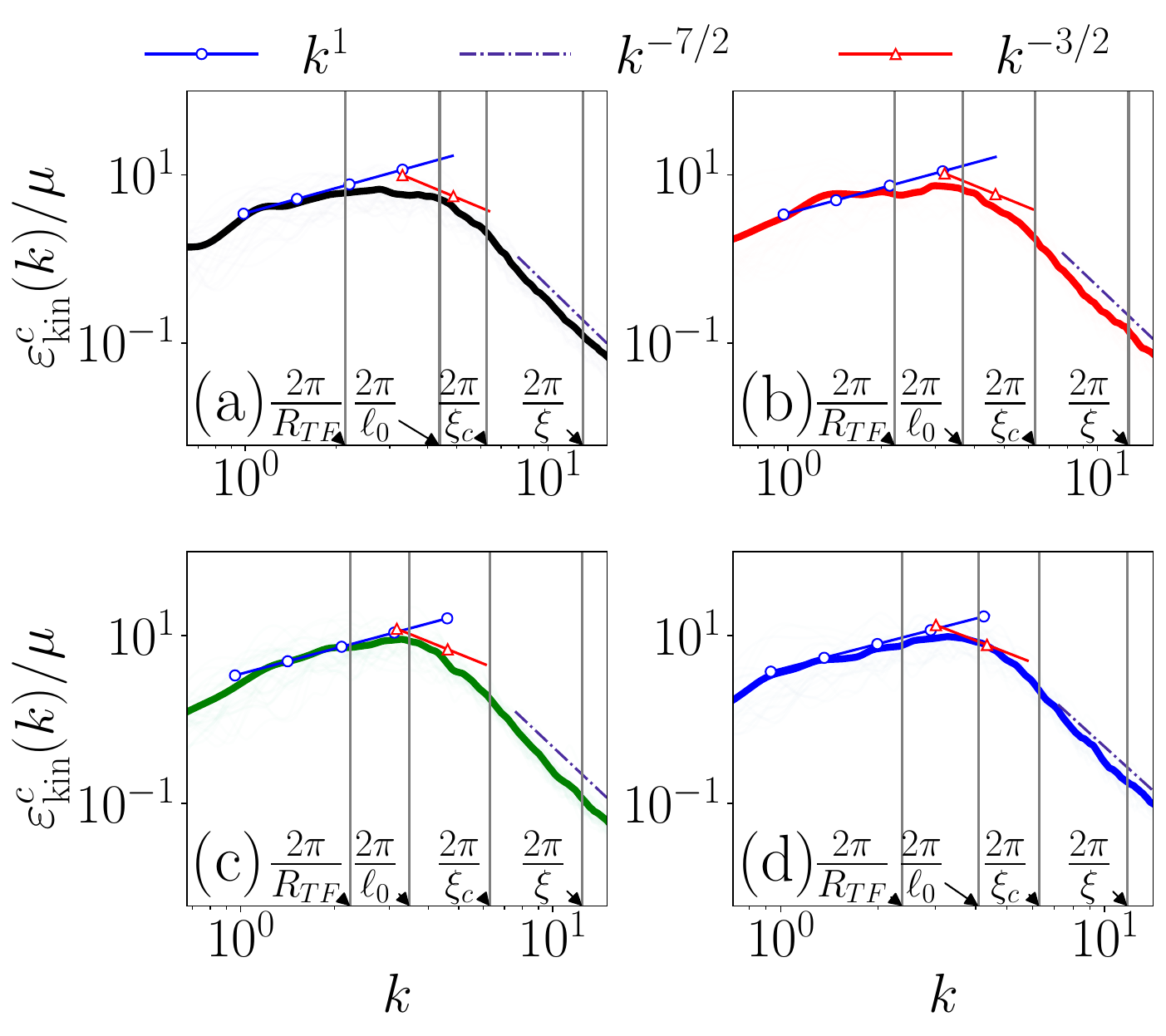}
    \caption{Plots depicting the compressible kinetic energy spectra for two condensates with circulation \(s=1\) at different collision velocities: (a) \(v = 6 \times 10^{-4}c\), with the spectral average over the time range \(t = 51\) to \(t = 320\) (in \(10^6\) years); (b) \(v = 7 \times 10^{-4}c\), in the time range \(t = 51\) to \(t = 320\); (c) \(v = 8 \times 10^{-4}c\), in the time range \(t = 128\) to \(t = 320\); and (d) \(v = 9 \times 10^{-4}c\), in the time range \(t = 128\) to \(t = 320\). All cases exhibit \(k\)-scaling behavior at smaller values of \(k\), transitioning to \(k^{-7/2}\) and \(k^{-3/2}\) scaling at larger values of \(k\).}
    \label{fig:ckinspectra3d_s11}
\end{figure}

In contrast, at larger wave numbers, the effects of compressible and incompressible dynamics become more comparable. Consequently, the scaling behavior is not solely dependent on the \(k^{-3}\) characteristic of the incompressible spectra; it also incorporates the \(k^{-5/2}\) behavior of the compressible counterpart, leading to a power law of \(k^{-17/2}\).  The compressible kinetic energy spectra (refer to Fig.~\ref{fig:ckinspectra3d_s11}) reveal a compelling insight into the intricate quantum turbulence within the system. The pronounced scaling observed at smaller \( k \) values is a clear indication of thermalization driven by density waves. This notable scaling predominantly manifests at the outer periphery or beyond the condensate, emphasizing the significant phase fluctuations in these regions. Furthermore, these fluctuations extend within the condensate radius, as evidenced by the observed \( k \)-scaling for \( k > 2\pi/R_{\mathrm{TF}} \) during rapid collisions [see Figs.~\ref{fig:ckinspectra3d_s11}(c)-(d)].

For slower collisions, a \(k^{-3/2}\) scaling is evident, suggesting the presence of weak-wave turbulence within the range \(2\pi/\ell_0 < k < 2\pi/\xi_c\). This \(k^{-3/2}\) scaling is comparatively weaker during faster collisions, where the \(k\) scaling becomes more dominant.
Notably, for faster collisions, the average spectra are computed starting at later times (from \(t \sim 128 \times 10^6\) years). This adjustment accounts for the large initial momentum, as the condensates initially pass through each other before colliding and forming a single condensate, as indicated by the secondary, larger peak in the gravitational wave luminosity profile shown in Fig.~\ref{fig:gw3d_s11}.

\section{Summary and conclusion}
\label{sec:conclusion}

We investigated the collision dynamics of two self-gravitating BECs, each possessing a circulation of $s=1$, with velocities sufficient to merge into a single condensate. The post-collision behavior exhibits similarities and deviations compared to phenomena observed in standard atomic BECs. Shortly after the collision, we observed interference fringes, which tend to drive the condensates apart. However, due to the selected velocity, the condensates lack the momentum to separate entirely and instead, they merge.
This merger leads to oscillatory motion, with the condensates repeatedly overlapping before settling into a single one. These oscillations are more pronounced and last longer with faster collisions. Eventually, the merged condensate reaches a circulation of \( s = 2 \), a configuration that is energetically unstable \cite{Nikolaieva2021}. In overextended timescales, the vortex structure collapses towards the periphery of the condensate, as seen in slower collisions where oscillations diminish rapidly. In contrast, for faster collisions (at \( v = 9 \times 10^{-4}c \)), the oscillations persist for timescales comparable to the age of the universe (approximately \( 13 \times 10^9 \) years), maintaining the central vortex structure. The collapse is also evident in the temporal profiles of the 3-D condensate radius. After the collision, the radius remains stable for an extended period before beginning to increase, which indicates the onset of vortex collapse that occurs earlier for slower collisions. The condensate oscillations, having a period of millions of years, can also be observed in the gravitational wave profiles and the kinetic energy components.

The interference fringes produced during the collision evolve into dark soliton rings \cite{Nikolaieva2023}, which decay into vortex-antivortex pairs via the snake instability \cite{Verma2017, sivakumar2024dynamic}. These vortices render the system energetically unstable, eventually collapsing to the condensate periphery. However, the central vortex core of circulation $s=2$ remains stable due to the merging \( s=1 \) vortices from the initially separated condensates.

We observed that the merged condensate exhibits turbulent behavior immediately after the collision similar to that reported in atomic BECs \cite{sivakumar2024dynamic, middleton2023}, where oscillations of the harmonic trap lead to a regime known as strong quantum turbulence \cite{barenghi2023types}. We have computed the incompressible kinetic energy spectra, which reveal a \(k^{-5/3}\) power-law scaling, indicative of a Kolmogorov cascade, at length scales around the intervortex distance \( \ell_0 \). This \(k^{-5/3}\) scaling ceases near the critical vortex core size \( \xi_c \), the upper limit for energetically stable vortices.

The decay mechanism of turbulence has also been analyzed by examining the temporal profiles of kinetic energy densities and their energy exchanges. During the turbulent stages, the incompressible kinetic energy is predominant, overshadowing both the compressible and quantum pressure components. However, in the later stages, the quantum pressure energy dominates over the energies associated with the condensate flow. This shift in dominance is attributed to the conversions of incompressible and compressible kinetic energies into the quantum pressure component. Since quantum pressure is a measure of the stability of the quantum fluid, this trend suggests that the condensate is striving to stabilize itself by reducing the condensate flow due to trap deformations.

Our research could provide valuable insights into the potential turbulent flows that emerge immediately following collisions between dark matter halos. The stability of these halos is critical, as, despite the significant initial momentum from the collisions, the resulting halo does not exhibit sustained flow dynamics. We observe several vortices with core sizes smaller than the critical core size, \(\xi_c\), which are exclusively generated by soliton decays during halo collisions. However, since their total circulation exceeds \(s=1\), these vortices are expelled toward the periphery of the halo. This phenomenon aligns with the findings of Ref. \cite{Nikolaieva2021} and has important implications for studying galactic rotations\cite{Castellanos2020}.

While vortices with \(s=2\) circulation have been shown to collapse on timescales comparable to the lifetime of the universe in dark-matter halos, we observe that halos with \(s=2\) circulation, formed via faster collisions of \(s=1\) condensates, persist beyond the lifetime of the universe. In Bose-Einstein condensate (BEC) experiments, as an alternative to harmonic confinement, such systems can be studied using softer-walled traps, which allow deformation by the condensate.

This work may be crucial for understanding the merger of binary neutron stars, such as the event GW170817, detected by the LIGO/Virgo gravitational wave observatories \cite{abbott2017}. According to Refs. \cite{hujeirat2020remnant, hujeirat2019glitching, hujeirat2018glitches}, pulsars are born with embryonic cores composed of incompressible superconducting supranuclear dense superfluids, the so-called SuSu-cores, surrounded by compressible and dissipative normal matter. Due to their static configurations and low-energy content, these cores can be considered as BECs, gravitationally confined by the surrounding shells.  %
\begin{figure}[ht!]
    \centering
    \includegraphics[width=0.99\linewidth]{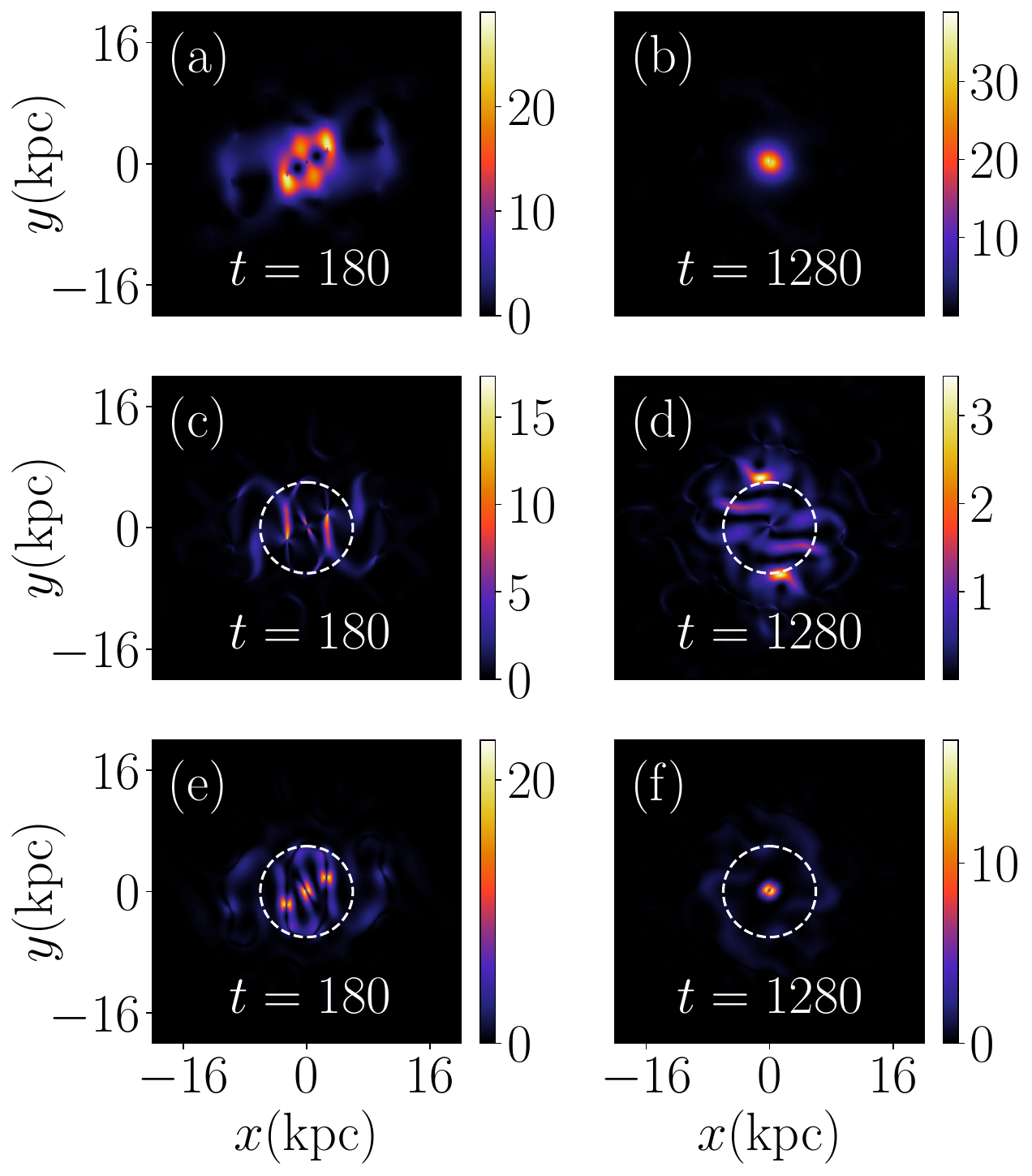}
    \caption{Energy densities in the \(x\)-\(y\) plane for (a)–(b) compressible (c)–(d) incompressible and (e)-(f) quantum pressure kinetic energies in the \(v = 9 \times 10^{-4}c\) case. The time \(t\) is expressed in units of \(10^6\) years, and the energy is scaled by \(\epsilon\) as described in Sec.~\ref{sec:intro}. The dashed white circle indicates the Thomas-Fermi radius of the condensate.}
    \label{fig:compdenxy_s11}
\end{figure}%

During the ring-down phase of the merger, the cores are predicted to merge smoothly and quietly, with a negligible loss of energy to the ambient media. However, a certain amount of rotational energy should be promptly ejected into the surroundings via the glitch phenomenon, which is observed to be associated with the cosmic evolution of pulsars. The ejection of rotational energy, in the form of vortices, would most likely give rise to turbulence in the boundary layer between the cores and the overlying shells of normal matter that exhibit the spectra reported in the present work. However, further investigations with sophisticated physics are still needed.
%During the ring-down of the merger, the cores are predicted to merge smoothly and quietly with a negligible loss of energy to the ambient media. Yet, a certain amount of rotational energy should be promptly ejected into the surroundings via the glitch phenomenon observed to associate the cosmic evolutions of pulsars. The ejection of the rotational energy in the form of vortices would most likely give rise to turbulent generation in the boundary layer between the cores and the overlying shells of normal matter that exhibit the spectra reported in the present work, though thorough investigations with sophisticated physics still need to be done.

\acknowledgments
The authors acknowledge the helpful discussions with Bibhas Ranjan Majhi. The work of A.S. and P.M. is supported by the MoE RUSA 2.0 (Bharathidasan University -- Physical Sciences).

\appendix

\section{Kinetic energy distribution in real space}
\label{appendix:a}
In the appendix, we present the spatial distribution of the incompressible and compressible components of the kinetic energy at different stages of the collision between the condensates. The incompressible and compressible energy densities for the $v=9\times 10^{-4}c$ case, shown in Fig.~\ref{fig:compdenxy_s11}(a)], are concentrated in certain regions. In contrast, in the long-term regime, the compressible energy settles and aggregates primarily in the central parts of the condensate. This behavior is markedly different from the evolution of compressible energy densities observed in atomic Bose-Einstein condensates (BECs), where the compressible density is more granular, with the average clump size being much smaller~\cite{sivakumar2024dynamic}. 
The aggregation of compressible energy into large clumps or smooth distributions suggests a lack of density waves in the long-term dynamics. The incompressible densities, which indicate the presence of vortices, are predominantly located within the condensate, as delineated by the Thomas-Fermi radius (white-dashed line) in Figs.~\ref{fig:compdenxy_s11}(c)-(d). However, in the long-term regime, the incompressible energy spreads primarily outside the Thomas-Fermi radius, indicating the expulsion of vortices to the condensate's periphery, except for a central vortex with $s=2$, which eventually collapses at later times. As quantum pressure arises due to sharp changes in density, the distribution of quantum pressure exhibits some sort of nucleation around the edges of the vortex cores as depicted in Figs.~\ref{fig:compdenxy_s11}(e)-(f).

%\bibliography{ref}
%\end{document}

%

\end{document}